\documentclass[11pt]{article}
\usepackage[latin1]{inputenc}
\usepackage[english]{babel}
\usepackage{graphicx}
\usepackage[namelimits]{amsmath}
\usepackage{amssymb}
\usepackage{amsmath}
\usepackage{amsthm}

\begin{document}

\title{Black Hole Entropy and Dilatations}

\author{Viaggiu Stefano
\\
Dipartimento di Matematica,\\ 
Universit\'a di Roma ``Tor Vergata'',\\
Via della Ricerca Scientifica, 1, I-00133 Roma, Italy\\
E-mail: viaggiu@mat.uniroma2.it\\
(or: stefano.viaggiu@ax0rm1.roma1.infn.it)}
\date{\today}\maketitle

\begin{abstract}
Dilatations by means of a constant factor 
can be seen in a double
way: as a simple change of units length or as a conformal
mapping of the starting spacetime into 
a ``stretched'' one with the same units length.
The numerical value of the black hole entropy  
depends on the interpretations made for 
the stretched manifold.
Further, we study the possibility to choose an unusual
``mass dependent'' normalization for the timelike Killing
vector for a Kerr black hole with and without a cosmic string.
\end{abstract}
\section{Introduction}
\noindent What does it happen to the black hole entropy
when we perform a constant global stretching of the 
manifold by means of a conformal constant factor $\Omega$, i.e.
\begin{equation}
d{s^{\prime}}^2={\Omega}^2 ds^2,\;\;\Omega >0\;?
\label{2}
\end{equation}
As a first
step we consider the Kerr solution written in Boyer-Lindquist
coordinates (see \cite{BL}):
\begin{eqnarray}
& &ds^2=\Sigma\left(d{\theta}^2+\frac{dr^2}{\Delta}\right)+
(r^2+a^2){\sin}^2\theta d{\phi}^2-dt^2+\nonumber\\
& &+\frac{2mr}{\Sigma}{(dt+a{\sin}^2\theta d\phi)}^2,
\;\Sigma=r^2+a^2{\cos}^2\theta,\;\Delta=r^2+a^2-2mr, \label{lal}
\end{eqnarray}
where $m$ and $a$ are respectively the mass and the spin
per unit mass of the source.
For $a^2<m^2$ the solution (\ref{lal}) describes a black hole
with outer horizon at $r_{+}=m+\sqrt{m^2-a^2}$. 
In this paper we use geometrized units with $G=c=\hbar=1$.
Hence mass, time and energy are measured in units of
length $L$. In these units the Newtonian constant $G$ and the light
velocity $c$ are dimensionless quantities and the fundamental 
constant $\hbar$ has dimension $L^2$. The black
hole entropy $S$ is given by the well known
Bekenstein-Hawking formula \cite{H1,Bek} $S=\frac{A}{4{l}_{p}^2}$
where $A$ is the horizon surface area given by
\begin{equation}
A={\int}_{r_+}\sqrt{g_{\theta\theta}}\sqrt{g_{\phi\phi}}d\phi
\label{cotoletta}
\end{equation}
and $l_p=\sqrt{\frac{G\hbar}{c^3}}$ is the Planck length that
in geometrized units has value $l_p=1$.
For the metric (\ref{lal}) we find
\begin{equation}
S=2\pi m(m+\sqrt{m^2-a^2}).\label{en}
\end{equation}
Besides, for the spacetime (\ref{lal}), we define the black hole
mass $m$ with respect to the timelike
Killing field 
${\xi}^{\nu}={(\frac{\partial}{\partial t})}^{\nu}$, where
${\xi}^{\nu}{\xi}_{\nu}=g_{tt}=\frac{a^2{\sin}^2\theta-\Delta}{\Sigma}$.
Furthermore, ``away from each mass source'', the metric can be recast,
in Cartesian coordinates, in the standard form
$ds^2=dx^2+dy^2+dz^2-dt^2$. If we have an asymptotically
flat spacetime, the normalization of ${\xi}^{\nu}$ 
at spatial infinity for the
spacetime (\ref{lal}) can be defined in the usual way
\begin{equation}
{({\xi}^{\nu}{\xi}_{\nu})}_{\infty}=-1.
\label{Tot}
\end{equation} 
We now consider 
the Kerr solution stretched by a factor ${\Omega}^2$, i.e.
\begin{eqnarray}
& &d{s^{\prime}}^2={\Omega}^2(r^2+a^2{\cos}^2\theta)
\left(d{\theta}^2+\frac{dr^2}
{r^2+a^2-2mr}\right)-
{\Omega}^2 dt^2+\nonumber\\
& &+{\Omega}^2(r^2+a^2){\sin}^2\theta
d{\phi}^2+
\frac{2mr{\Omega}^2}
{(r^2+a^2{\cos}^2\theta)}
{[dt+a{\sin}^2\theta d{\phi}]}^2.\label{tilt}
\end{eqnarray}
What is the relation between the spacetimes
(\ref{lal}) and (\ref{tilt})?
This paper tries to overcome this question.\\
In section 2 we study the transformation (\ref{2}) in
relation to the entropy. In section 3 we discuss the normalization
of the timelike Killing field. In section 4 the same
discussion is made for a particular simple non asymptotically
flat case: the Kerr black hole with a static cosmic
string. 
\section{Dilatations in asymptotically flat spacetimes}
\indent As a first consideration note that $\Omega$
is dimensionless. Further, the Einstein's tensor
$G_{\mu\nu}=R_{\mu\nu}-\frac{g_{\mu\nu}}{2}R$ is invariant
under (\ref{2}). Thus, the transformation (\ref{2}) has only
two possible interpretations:\\
{\bf First interpretation}: {\it passive point of view}.\\
The transformation (\ref{2}) is a simple change in
units length. In other words, if in (\ref{lal}) 
we measure the length in meters and if $\Omega=10^2$, 
then we measure
time, mass and energy in (\ref{tilt}) in centimeters. If we take
$r^{\prime}=\Omega r, t^{\prime}=\Omega t, m^{\prime}=\Omega m,
a^{\prime}=\Omega a$, the metric (\ref{tilt}) becomes
\begin{eqnarray}
& &d{s^{\prime}}^2=({r^{\prime}}^2+{a^{\prime}}^2{\cos}^2\theta)
\left(d{\theta}^2+\frac{d{r^{\prime}}^2}
{{r^{\prime}}^2+{a^{\prime}}^2-
2m^{\prime}r^{\prime}}\right)-d{t^{\prime}}^2+\nonumber\\
& &+({r^{\prime}}^2+{a^{\prime}}^2){\sin}^2\theta d{\phi}^2+
\frac{2m^{\prime}r^{\prime}}{{(r^{\prime}}^2+{a^{\prime}}^2{\cos}^2\theta)}
{\left[dt^{\prime}+a^{\prime}{\sin}^2\theta d{\phi}\right]}^2. \label{segr}
\end{eqnarray} 
The entropy formula $S^{\prime}$ for the line element (\ref{segr}) gives
$S^{\prime}=S$ because $A^{\prime}={\Omega}^2 A$
and ${l}_{p}^{\prime}={\Omega}$.
This means that the microscopic realizations
of the black hole (\ref{lal}) are independent of the 
units used.\\
{\bf Second interpretation}: {\it active point of view}.\\
The transformation (\ref{2}) is a conformal mapping of the
spacetime (\ref{lal}) into a ``stretched'' spacetime. 
Obviously, in this second case,
$G=c=1$ and also $\hbar=1$ (${l}_{p}^{\prime}=l_p=1$) because
the units length are unchanged. Therefore, the entropy formula gives
$S^{\prime}={\Omega}^2 S$ and the counting of the microscopic states 
fails to be the same for spacetimes (\ref{lal}) and (\ref{tilt}).\\[5pt]
\indent On the grounds of these
interpretations, the entropy value depends on the {\it active} and 
{\it passive}
point of view.
Further, the weak limit of Einstein's equations leads to 
the Poissonian equation
${\nabla}^2 U=4\pi\rho$, where $U$ is the 
Newtonian potential and
$\rho$ is the mass density of the source \cite{Fe}. The only 
change caused by (\ref{2}) in 
the weak limit is given by $U\rightarrow U_{\Omega}=U-ln\Omega$, 
that is a simple
gauge transformation which cannot affect Poissonian equation.\\
This can be of some interest for the so called 
Immirzi ambiguity \cite{Im1,Im2},
that arises in the context of loop gravity \cite{Lo,AS} 
by means of the Ashtekar 
formulation of general relativity, or for considerations 
involving a generic quantum
gravitational theory \cite{Birr}. In loop gravity the 
counting of quantum states for the
black hole entropy $S$ is affected by an arbitrary 
multiplicative constant $\beta$ that
cannot be fixed a priori by the quantization procedure 
\cite{Rov} and by the low
energy physics.\\
When we have a generic quantum theory, 
it is not a trivial question to ponder, in relation
to the {\it active} point of view,
which is between $ds^2$ and ${\Omega}^2 ds^2$ the  
classical reference spacetime of the quantum
theory. In this 
context, it is expected that in loop quantum 
gravity the entropy is affected by an arbitrary
multiplicative constant. In literature the analogy 
between scale transformations and the
Immirzi ambiguity has been taken in account in \cite{M1}, 
but only in the case of
the {\it passive} point of view.\\
\indent Now, let us suppose that $\Omega$ is
a function of the parameters characterizing the black hole,
i.e. $\Omega=F(m,a)$, with the normalization (\ref{Tot}).
Adopting the {\it passive} point of view 
, the entropy $S^{\prime}$ for the 
spacetime (\ref{tilt}) is
\begin{equation}
S^{\prime}=\frac{2\pi m^{\prime}(m^{\prime}+
\sqrt{m^{\prime 2}-a^{\prime 2}})}{F^2},
\label{ende}
\end{equation}
where $m^{\prime}=Fm, a^{\prime}=Fa$.
Conversely, adopting the {\it active} point of view, 
the entropy $S^{\prime}$ becomes
\begin{equation}
S^{\prime}=2\pi m^{\prime}(m^{\prime}+
\sqrt{m^{\prime 2}-a^{\prime 2}}).
\label{ended}
\end{equation} 
What does it happen if we choose
another normalization for the Killing vector?
In the next sections we will study this problem from
the {\it active} point of view. The {\it active} point
of view is the most interesting case because 
all the observers use the same units length, independently on
the scale transformation (\ref{2}).  
\section{Normalization of the timelike Killing vector for asymptotically
flat spacetimes} 
\indent  In this
section  we use another normalization instead of (\ref{Tot}) 
as, for example, for black holes in 2D dilaton
gravity \cite{Fa}. In this case, the usual
normalization (\ref{Tot}) does not lead to the energy
conservation: to achieve energy conservation is necessary
a ``mass dependent'' normalization.
Following this point of view, we could use the 
normalization
\begin{equation}
{({\xi}^{\nu}{\xi}_{\nu})}_{\infty}=
-{\Omega}^2=-F^{2}(m, a).
\label{lea}
\end{equation}
With respect to
(\ref{lea}) we define the mass of the black hole.
In this case, if we ``see''
the spacetime (\ref{lal}) with the normalization (\ref{lea}),
the parameters which are effectively measured are not $m$ and $a$,
but rather the rescaled ones $m^{\prime}, a^{\prime}$,
where $m=m^{\prime}F,
a=a^{\prime}F$.
Note that, since $\frac{a}{m}=\frac{a^{\prime}}{m^{\prime}}$, if we
take $F=F(a/m)$ then $F(a/m)=F(a^{\prime}/m^{\prime})$.
This fails if 
the function $F$ depends on $\hbar$. For example, if
$F=ma$ ($\hbar=1$ in our units ) then, 
in terms of $m^{\prime}, a^{\prime}$,
$F=\frac{1}{m^{\prime}a^{\prime}}$. 
This lack of symmetry when $\hbar$ is present in $F$ has
an interesting analogy with the cosmological constant $\Lambda$.
In fact, if $\Lambda$ is present in Einstein's equations, i.e.
${G}_{\mu\nu}\rightarrow {G}_{\mu\nu}+\Lambda g_{\mu\nu}$, then 
the invariance of ${G}_{\mu\nu}$ under (\ref{2}) is broken.
To regain invariance we must rescale $\Lambda$ as
$\Lambda\rightarrow\frac{\Lambda}{{\Omega}^2}$. Remember that it
is the presence of $\hbar$ in the formula for the entropy $S$ that 
breaks the symmetry between the {\it active} and the {\it passive}
point of view. Besides, the entropy of a black hole arises when quantum
reasonings are taken in account (classically the
black hole entropy is exactly zero) and the presence of
$\hbar$ is an indication of such quantum reasonings. By analogy,
this suggests that
the origin of $\Lambda$ could be found in quantum theory.\\  
Generally, the increasing
or decreasing character of the entropy is related to the
normalization chosen for ${\xi}^{\nu}$.
In practice, it is the presence of the black hole 
that can justify the normalization
(\ref{lea}). Conversely, if we have a Minkowskian spacetime,
no objects present in the universe modify the flat geometry.
Therefore, we have no grounds to choose an
``objects dependent'' normalization. In other words,
in a Minkowskian spacetime, surfaces and volumes are 
``absolute'' objects that do not depend on the 
state of the matter and thus we can define a
``privileged reference'' metric. 
The function $F$ presents in  (\ref{lea}) is arbitrary:
by varying the function $F$ we choose the scale at which we see a given 
spacetime. 
For example, in \cite{Ka} one compute the 
logarithmic correction to the Bekenstein-Hawking formula in the formulation
of ``quantum geometry'' of Ashtekar \cite{As2} and in certain string theories
\cite{Car}.
The modified entropy is 
\begin{equation}
S=\frac{A}{4}-\frac{3}{2}ln\frac{A}{4}+\cdots
\label{indi}
\end{equation}
If we ``see'' the spacetime (\ref{lal}) with the normalization (\ref{lea}),
then we find the scale at which the corrections arise: we must impose
\begin{eqnarray}
S&=&2\pi m(m+\sqrt{m^2-a^2})=
2\pi m^{\prime}(m^{\prime}+\sqrt{m^{\prime 2}-a^{\prime 2}})F^2=\nonumber\\
&=&2\pi m^{\prime}(m^{\prime}+\sqrt{m^{\prime 2}-a^{\prime 2}})-\nonumber\\
&-&\frac{3}{2}ln[2\pi m^{\prime}(m^{\prime}+
\sqrt{m^{\prime 2}-a^{\prime 2}})]+\cdots\label{Ale}
\end{eqnarray}
Hence, by posing $F^2=\zeta$, we find
\begin{equation}
ln\frac{\zeta}{S} =\frac{2}{3}S-\frac{2}{3}S\frac{1}{\zeta}.
\label{Sari}
\end{equation} 
Graphically, it is easy to see that equation (\ref{Sari}) has
not solutions for $0<S<\frac{3}{2}-\frac{3}{2}ln\frac{3}{2}$,
one solution for $S=\frac{3}{2}-\frac{3}{2}ln\frac{3}{2}$ and two solutions
for $S>\frac{3}{2}-\frac{3}{2}ln\frac{3}{2}$. When $S<1$
we have microscopic black holes with mass $m<1$ 
expressed in terms of the unit Planck length $l_p$. 
For black holes with mass larger than the Planck length we have two 
solutions, the first
with $F>1$ and the latter with $F<1$. 
The leading correction to the
entropy formula has been obtained by taking the ``mass dependent'' 
normalization (\ref{lea}). In other words, when the 
entropy for the spacetime (\ref{lal}), at the scale given by (\ref{lea}),
is expressed in terms of the parameters measured, i.e.
$m^{\prime}, a^{\prime}$, corrections arise, provided that the 
equation (\ref{Sari}) is satisfied.  
\section{Non asymptotically flat case}
\indent We consider now the case of a non asymptotically
flat spacetime by ``preserving'' the
{\it active} point of view. It is a well known fact
\cite{Wil} that, in polar coordinates, the spacetime
\begin{equation}
d{{s}^{\prime}}^2=d{\rho}^2+dz^2+
{(1-4\mu)}^2{\rho}^2 d{\phi}^2-dt^2,
\label{cicc}
\end{equation}
when $0<\mu<\frac{1}{4}$, is a solution of Einstein equations
$G_{\mu\nu}=8\pi T_{\mu\nu}$ with 
$T_{\mu\nu}=\mu\delta (x^{\prime})\delta (y^{\prime})diag(1,0,0,-1)$.
The solution (\ref{cicc}) represents a static string (cosmic string) with
a mass distribution on the $z$ axis, where $\mu$ is the mass density
of the source. The parameter $B=1-4\mu$ gives the topological defect 
(angle deficit) of
the spacetime. It is also known \cite{GV} that in the limit
${\rho}\rightarrow 0$ the quantity
\begin{equation}
\Delta\Phi({\rho})=2\pi-\frac{{\int}^{2\pi}_{0}
\sqrt{g_{\phi\phi}} d\phi}
{{\int}^{\rho}_{0}
\sqrt{g_{\rho\rho}} d{\rho}}
\label{in}
\end{equation}
is directly related to the 
energy density per unit length of the string ($\Delta\Phi(0)=8\pi\mu$).
If $\Delta\Phi(0)=0$ the topological defect disappears. For the metric
(\ref{cicc}) we get $\Delta\Phi(0)=2\pi(1-B)$ ($B<1$).
The spacetime (\ref{cicc}) is locally, but not globally, Minkowskian.
Now, also the metric
\begin{equation}
ds^2=\frac{1}{{(1-4\mu)}^2}(d{\rho}^2+dz^2)+{\rho}^2d{\phi}^2-\frac{dt^2}
{{(1-4{\mu})}^2}
\label{Bar}
\end{equation}
is a solution of the same equations satisfied by 
the line element (\ref{cicc}). 
Both solutions (\ref{cicc}) and (\ref{Bar}) are invariant under Lorentz
boosts along $z$ axis.
These two  solutions
 are joined by (\ref{2}) with ${\Omega}^2=B^2$, 
and  the 
parameter $\mu$ in both solutions  
is the mass density of the string source. This can also be understood 
from expression (\ref{in}) invariant under a constant stretching
of the line element. In fact the spacetimes (\ref{cicc}) 
and (\ref{Bar}) have the same 
physical interpretation and describe the same spacetime
``seen'' at different scales. Note that if we choose for $B$ 
a value different from $B=1-4\mu$, for example 
$\tilde{B}=\frac{1+{\tilde{\mu}}^2}{1+4\tilde{\mu}}$
($\tilde{B}<1\rightarrow 0<\tilde{\mu}<\frac{1}{4})$, the parameter
$\tilde{\mu}$ cannot be interpreted as the mass density of the string
because we must always have $\Delta\Phi(0)=2\pi(1-\tilde{B})=8\pi\mu$,
where $\mu$ is the true mass density of the string.
In practice, the parameter $\mu$ is dimensionless in our units and
thus it is a scale invariant object under (\ref{2}).
In any case both metrics (\ref{cicc}) and (\ref{Bar}) are locally but
not globally equivalent to the Minkowskian spacetime and both describe
a static string with mass density $\mu$ on the $z$ axis.\\
\indent A rotating black hole with a cosmic string of mass density
$\mu$ along $z$ axis \cite{Al,Gal,Viag}
in Boyer-Lindquist
, according to the
asymptotic form (\ref{Bar}) with the same notation of 
equation (\ref{lal}), has the line element
\begin{eqnarray}
ds^2&=&\frac{\Sigma}{B^2}\left(d{\theta}^2+\frac{dr^2}{\Delta}\right)+
(r^2+a^2){\sin}^2\theta d{\phi}^2-\frac{dt^2}{B^2}+\nonumber\\
&+&\frac{2mr}{\Sigma}{\left(\frac{dt}{B}+a{\sin}^2\theta d\phi\right)}^2
.\label{gin}
\end{eqnarray}
Thanks to (\ref{in}) we obtain, for solution
(\ref{gin}),  
$\Delta\Phi(0)=2\pi(1-B)$.
The horizon surface area of black hole is  
$A=\frac{4\pi({r_{+}}^2+a^2)}{B}$ with $r_+=m+\sqrt{m^2-a^2}$.\\
If we take 
$g_{\mu\nu}\rightarrow g_{\mu\nu}B^2$ we have:
\begin{eqnarray}
& &d{s^{\prime}}^2={\Sigma}\left(d{\theta}^2+
\frac{dr^2}{\Delta}\right)
+B^2(r^2+a^2){\sin}^2\theta d{\phi}^2+\nonumber\\
& &+\frac{2 mr}{\Sigma}
{\left(dt+Ba{\sin}^2\theta d{\phi}\right)}^2
-dt^2. \label{cicala}
\end{eqnarray}
Formula (\ref{in}) gives again $\Delta\Phi(0)=2\pi(1-B)$ and
$r_{+}=m+\sqrt{m^2-a^2}$.\\
We can measure the mass density $\mu$ at $r=\infty$ and therefore, for 
the discussion above, the parameter $\mu$ 
is the mass density of the string for both
metrics (\ref{gin}) and (\ref{cicala}). 
We must choose a normalization for the 
timelike Killing vector with respect to which we define the black hole mass
$m$. Generally, if we have a non asymptotically flat spacetime there is not
an usual way to fix the normalization of ${\xi}^{\nu}$. 
In our case, we have to choose
one of the two asymptotic forms (\ref{cicc}) and (\ref{Bar}).
Let us assume that the asymptotic line element  (\ref{cicc}) is our
``reference'' metric. For the entropy $S^{\prime}$ we have:
\begin{equation}
S^{\prime}=2\pi m(m+\sqrt{m^2-a^2})(1-4\mu).
\label{old}
\end{equation}
The entropy formula (\ref{old}) is a {\bf decreasing} function of the
parameter $\mu$. Now, by keeping the normalization
(\ref{Tot}), we consider the solution (\ref{gin})
with $ds^{\prime 2}=B^2 ds^2$. The entropy for this spacetime is
\begin{eqnarray}
S(m,a,\mu)&=&\frac{2\pi m(m+\sqrt{m^2-a^2})}{(1-4\mu)}=\nonumber\\
&=&S(m^{\prime},a^{\prime},\mu)=2\pi m^{\prime}
(m^{\prime}+\sqrt{m^{\prime 2}-a^{\prime 2}})(1-4\mu)
\label{lil}
\end{eqnarray}
with $m=m^{\prime}B, a=a^{\prime}B$, where $m^{\prime}$ and
$a^{\prime}$ are respectively the mass and the spin density
effectively measured in the spacetime (\ref{gin}). 
Thus, if we consider the function $S$ for spacetime (\ref{gin}), with
the normalization (\ref{Tot}), as a function of $m^{\prime}, a^{\prime},
\mu$, the entropy continues to be a decreasing function of the 
parameter $\mu$. The objects physically relevant are the observable
quantities that, for spacetime (\ref{gin}) with the normalization
(\ref{Tot}), are $m^{\prime}$ and $a^{\prime}$.\\  
Note that, if we take the limit
$\mu\rightarrow\frac{1}{4}$, then $S^{\prime}\rightarrow 0$.
Moreover, in this limit, for the black hole temperature $T_{BH}$
with $T_{BH}=\frac{\partial m}{\partial S}$, we find that
$T_{BH}\rightarrow\infty$. In this limit the metric becomes singular.
Cosmic strings with string tension $\mu$ appear in the context of 
cosmological models in order to act as seeds for galaxy formation.
It is interesting to note that the very hot limit with high mass
density $\mu$ corresponds to the zero entropy limit.
This could mean that black holes may have been formed
during phase transition  when the
matter has been extremely hot and dense. i.e. at the very early stage 
of the universe.\\ 
If we  choose as asymptotical  ``reference''
metric the expression (\ref{Bar}), we can
take
\begin{equation}
{({\xi}^{\nu}{\xi}_{\nu})}_{\infty}=-B^{-2}.
\label{brig}
\end{equation} 
With this choice the parameters $m$ and $a$ that appear in the line element
(\ref{gin}) are now the physical one measured and
thus entropy is
\begin{equation}
S(m,a,\mu)=\frac{2\pi m(m+\sqrt{m^2-a^2})}{(1-4\mu)}
\label{Vac}
\end{equation}
that is an {\bf increasing} function of the parameter $\mu$.
For the 
spacetime (\ref{cicala}) the entropy, with the new normalization
(\ref{brig}), becomes
\begin{equation}
S^{\prime}(m^{\prime},a^{\prime},\mu)=
\frac{2\pi m^{\prime}(m^{\prime}+\sqrt{m^{\prime 2}-a^{\prime 2}})}
{(1-4\mu)}
\label{olg}
\end{equation}
with $a^{\prime}=Ba, m^{\prime}=Bm$. Now the parameters $m^{\prime}$
and $a^{\prime}$ are respectively the mass and spin density
of the black hole source measured
in (\ref{cicala}). Also in this case, the entropy is an increasing
function of the parameter $\mu$.\\
Generally, we can multiply
solution (\ref{gin}) by $B^{\gamma}$, where $\gamma$ is 
any real constant, and choose the normalization
of ${\xi}^{\nu}$ according to the asymptotic metric so obtained.
Therefore the black hole entropy 
with a cosmic string with respect to the ``reference'' metric so obtained 
becomes
\begin{equation}
S=2\pi m(m+\sqrt{m^2-a^2})B^{\gamma-1}\;\;,\;\;
{({\xi}^{\nu}{\xi}_{\nu})}_{\infty}=-B^{\gamma-2},
\label{exodus}
\end{equation}
If we take $\gamma =1$ in expression (\ref{exodus}), then   
the entropy formula is independent on the 
parameter $\mu$ . Besides, if $\gamma<1$ the high mass density limit
$\mu\rightarrow \frac{1}{4}$ leads to $S\rightarrow\infty$ and
$T_{BH}\rightarrow 0$, in contrast to the situation (\ref{old}).
The choice of normalization for ${\xi}^{\nu}$, i.e. the choice of the 
``reference'' metric in relation to the ``active'' point of view,
is equivalent to choose an arbitrary scale energy
with respect to which we ``see'' the other solutions joined by (\ref{2}).\\
Taking the phraseology of ordinary quantum field theory, one can think
at $m^{\prime}$ as a kind of ``interacting'' mass and at $m$ as a ``bare'' 
one. In fact, the parameters present in the action of a ordinary
quantum fields theory, as it happens in the standard electroweak interaction
model, are not the ones measured in the effective theory:
it is by means of the {\it renormalization} 
procedure that one can define  the 
``interacting'' parameters really observed; the 
``bare'', non ``interacting'', parameters are meaningless from
a physical point of view. 
In the {\it renormalization group} approach the mass is a physical
parameter that depends on the scale under consideration.\\
Finally, the imposition 
${({\xi}^{\nu}{\xi}_{\nu})}_{\infty}=-B^{\gamma-2}$
with $m^{\prime}=mB^{(1-\frac{\gamma}{2})}$
is not bizarre because the parameter $\mu$ appears also in the
asymptotic line element
and thus we can choose a scale, with respect to which we measure the
mass $m^{\prime}$, which takes into account such presence at spatial
infinity.
In fact, according to Mach (see \cite{Mach,Einst})
, the presence of the string can modify the inertia and the
gravitational mass of a body. In this sense the presence of the string
can justify a kind of ``interacting'' normalization for
${\xi}^{\nu}$ depending on the parameter $\mu$.

\end{document}